\documentclass[11pt]{article}
\pdfoutput=1

\usepackage{euscript}
\usepackage{amssymb}
\usepackage{amsfonts}
\usepackage{amsbsy}
\usepackage{epsfig}
\usepackage{amsthm}
\usepackage{amscd}
\usepackage{amstext}
\usepackage{verbatim}
\usepackage{amsmath}
\usepackage{cancel}
\usepackage{capt-of}
\usepackage{empheq}
\usepackage{subfigure}
\usepackage{xcolor}

\usepackage{cite}
\usepackage{bm}
\usepackage{authblk}
\usepackage[T1]{fontenc}

\numberwithin{equation}{section}

\usepackage[pdftex,linktoc=all]{hyperref}
\hypersetup{ 
colorlinks=true, 
linkcolor=black, 
citecolor=red, 
}

\def\ben{\begin{equation}}
\def\een{\end{equation}}
\def\half{{\textstyle{\frac{1}{2}}}}
\let\a=\alpha  \let\g=\gamma  
   
     \let\r=\rho

\let\pa=\partial
\def\be{\begin{equation}}
\def\ee{\end{equation}}
\def\beq{\begin{equation}}
\def\eeq{\end{equation}}
\def\ba{\begin{array}}
\def\ea{\end{array}}

\def\dalemb#1#2{{\vbox{\hrule height .#2pt
       \hbox{\vrule width.#2pt height#1pt \kern#1pt
               \vrule width.#2pt}
       \hrule height.#2pt}}}

\newcommand{\bea}{\begin{eqnarray}}
\newcommand{\eea}{\end{eqnarray}}

\def\vep{{\varepsilon}}

\makeatletter
\newcommand*\bigcdot{\mathpalette\bigcdot@{.5}}
\newcommand*\bigcdot@[2]{\mathbin{\vcenter{\hbox{\scalebox{#2}{$\m@th#1\bullet$}}}}}
\makeatother

\renewcommand{\eqref}[1]{(\ref{#1})}

\def\Lag{{\mathcal{L}}}

\def\ocal{{\mathcal{O}}}

\title{Gravitational duals to the grand canonical ensemble \\ abhor Cauchy horizons}

\author{Sean A. Hartnoll$^1$, Gary T. Horowitz$^2$, Jorrit Kruthoff$^1$ and Jorge E. Santos$^{3,4}$
\\ {\it $^1$ Department of Physics, Stanford University, Stanford, CA 94305-4060, USA} \\

{\it $^2$ Department of Physics, University of California, Santa Barbara, CA 93106, USA} \\
{\it $^3$ DAMTP, University of Cambridge, Wilberforce Road, Cambridge CB3 0WA, UK} \\
{\it $^4$ Institute for Advanced Study, Princeton, NJ 08540, USA}
}

\begin{document}
\frenchspacing

\maketitle

\begin{abstract}

The gravitational dual to the grand canonical ensemble of a large $N$ holographic theory is a charged black hole. These spacetimes --- for example Reissner-Nordstr\"om-AdS ---  can have  Cauchy horizons that render the classical gravitational dynamics of the black hole interior incomplete. We show that a (spatially uniform) deformation of the CFT by a neutral scalar operator generically leads to a black hole with no inner horizon. There is instead a spacelike Kasner singularity in the interior. For relevant deformations, Cauchy horizons never form. For certain irrelevant deformations, Cauchy horizons can exist at one specific temperature. We show that the scalar field triggers a rapid collapse of the Einstein-Rosen bridge at the would-be Cauchy horizon. Finally, we make some observations on the interior of charged dilatonic black holes where the Kasner exponent at the singularity exhibits an attractor mechanism in the low temperature limit.

\end{abstract}

\newpage

 \tableofcontents

\section{Introduction}

Black hole interiors present many theoretical challenges, at both a classical and quantum level. One of these challenges is the singularity at which spacetime ends \cite{Hawking:1969sw}. The classical approach to generic singularities is expected to be very complicated \cite{BKL}, while the classical description itself eventually breaks down as curvatures become large. Another challenge is the possible presence of Cauchy horizons, at which the predictability of the classical dynamics breaks down, even away from regions with large curvature \cite{Geroch:1970uw}. The strong cosmic censorship conjecture posits that such Cauchy horizons are artifacts of some highly symmetric solutions that are known analytically, and do not arise from  generic initial data \cite{Penrose1979}.

In holographic duality, eternal black holes in asymptotically AdS spacetimes arise as thermofield double states in a large $N$ CFT \cite{Maldacena:2001kr}. This fact has led to rigorous boundary probes of the black hole interior using \emph{e.g.} entanglement entropy \cite{Hartman:2013qma}. So far, probes of the region close to spacelike singularities have required analytic continuation of boundary correlation functions \cite{Fidkowski:2003nf} and do not appear to directly access Cauchy horizons \cite{Brecher:2004gn}. Holographic arguments suggest that in general, Cauchy horizons do not survive in the full quantum gravity theory \cite{Papadodimas:2019msp, Balasubramanian:2019qwk}. (To first subleading order in large $N$, the three-dimensional BTZ black hole maintains its Cauchy horizon \cite{Dias:2019ery,Hollands:2019whz}, but it is probably destroyed at higher order \cite{Emparan:2020rnp}.)  With ongoing interest in probing the interior, it is important not to be led astray by aspects of the spacetime that may be artifacts of the simplest known solutions. The most studied solutions in holography are the Schwarzschild-AdS spacetime as dual to the canonical ensemble \cite{Witten:1998qj} and Reissner-Nordstr\"om-AdS (RN-AdS) spacetime as dual to the grand canonical ensemble \cite{Chamblin:1999tk}. These have rather particular singularity structures and RN-AdS has an inner Cauchy horizon.

While a fully generic interior will be highly inhomogeneous, a tractable step in the direction of genericity for uncharged black holes was considered in \cite{Frenkel:2020ysx}, motivated from the dual field theory perspective. The simplest AdS black holes spacetimes are dual to the thermofield double state of a CFT. The CFT itself is often non-generic within the space of field theories in the sense that relevant deformations (such as mass terms) must be tuned to zero to remain at the critical point.  To probe more generic thermal states, the relevant operators can be turned on. This can be done with a coupling constant that is uniform in the boundary spacetime. Relevant operators are described in the bulk by scalar fields with negative mass squared (but above the Breitenlohner-Freedman bound \cite{Breitenlohner:1982jf}). Sourcing such fields at the AdS boundary should be expected to produce more generic black hole solutions. In \cite{Frenkel:2020ysx} it was found that, indeed, these solutions had a more generic behavior at the black hole singularity, with the Schwarzschild singularity arising as a fine-tuned special case. The more generic behavior is described by a one parameter family of homogeneous, anisotropic cosmologies known as Kasner spacetimes. Thus, genericity at the boundary led to genericity at the singularity. In this paper we will ask an analogous and perhaps more consequential question for charged black holes: does turning on a relevant deformation of the boundary theory remove the Cauchy horizon? The answer will be that it does.\footnote{It was noted in \cite{Dias:2019ery} that a multi-trace deformation of a two-dimensional CFT can destroy the Cauchy horizon of a BTZ black hole.}

The boundary perspective motivates a holographic version of strong cosmic censorship with a slightly different flavor from the conventional one. Usually one asks about the stability of Cauchy horizons in the space of generic initial conditions. Holographically one can ask instead whether a generic time-independent thermal state of the boundary theory leads --- in the classical large $N$ limit --- to a dual black hole with a Cauchy horizon. As we have explained above, from this boundary perspective RN-AdS is not generic if the CFT has relevant deformations that have been fine tuned to zero. The results below are evidence in favor of such a notion of holographic strong cosmic censorship. 

Three comments should be made here. Firstly, since the radial black hole coordinate becomes timelike in the interior, what start off as asymptotic boundary conditions ultimately play the role of initial conditions for the interior. Thus the two formulations of strong cosmic censorship have some overlap. Secondly, a key step in attempting to prove strong cosmic censorship involves establishing that perturbations outside the horizon do not decay too quickly, so that they can build up inside and prevent the formation of a Cauchy horizon \cite{Kehle:2018zws}. For example, it has recently been shown that the Cauchy horizon is stable for some charged black holes in de Sitter space where the perturbations fall off exponentially fast outside the horizon \cite{Cardoso:2017soq,Dias:2018etb}. A source at the boundary that is present for all time clearly helps with this issue and therefore this holographic version is weaker than the conventional one. Thirdly, inhomogeneous deformations of the boundary can induce regions of strong curvature that are directly visible to boundary observers \cite{Horowitz:2016ezu,Crisford:2017zpi}. These are violations of weak cosmic censorship, which is not the focus of our present discussion. (The effect of a small inhomogeneous deformation on a Cauchy horizon is discussed in \cite{Maeda:2011pk}.)

We have focused on relevant deformations so far, but irrelevant deformations of the CFT will also be generically present at nonzero temperature if the CFT is obtained as the IR fixed point of some UV completion such as a lattice model. While relevant deformations always remove Cauchy horizons, we will show that certain irrelevant deformations, dual to a bulk scalar with $m^2 > 0$ (positive mass squared), do allow them. But these Cauchy horizons can only occur at a discrete set of $m^2$ for each temperature. 
Irrelevant deformations destroy the asymptotic AdS region, which must either be explicitly cut off or otherwise allowed to flow to some distinct UV fixed point where the operator is relevant. However, our discussion will only require knowledge of the spacetime inside the event horizon.

\begin{figure}[t]
\centering
\includegraphics[width=0.6\textwidth]{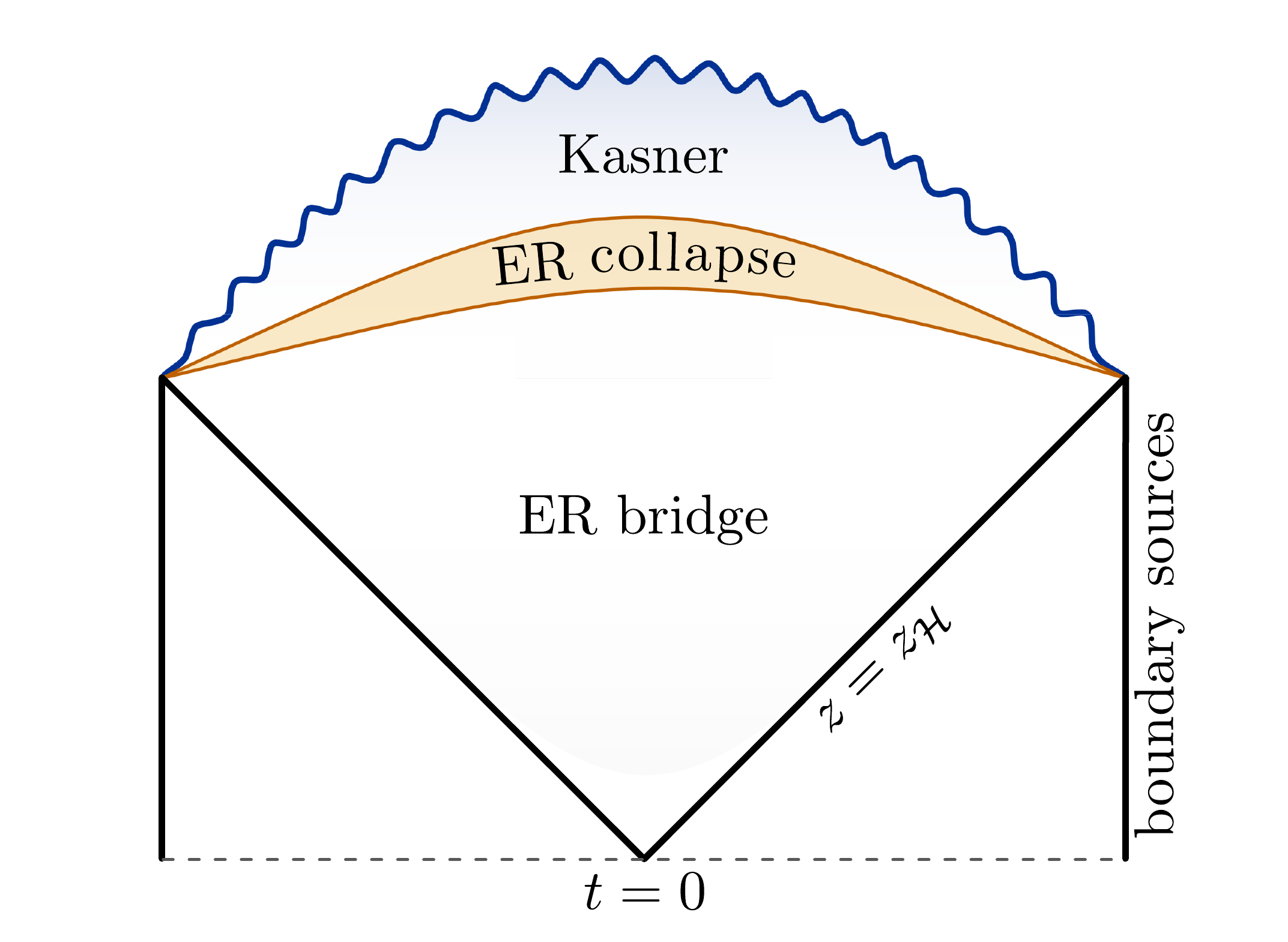}
\caption{Schematic illustration of the interior of a charged AdS black hole in Einstein-Maxwell theory with a small scalar field source. It begins close to RN-AdS with its standard Einstein-Rosen (ER) bridge. This undergoes a rapid collapse discussed in \S\ref{sec:collapse},  followed by a Kasner cosmology towards the singularity, discussed in
\S \ref{sec:Kasner}. The Penrose diagram is considered in more detail \S \ref{sec:Penrose}. The black solid lines indicate the boundary and horizon of the geometry.}
\label{fig:summary}
\end{figure}

The fate of the Cauchy horizon is especially dramatic for the case of a small deformation of the AdS boundary. The solution remains close to RN-AdS until one approaches the would-be Cauchy horizon. At that point there is a rapid collapse of the Einstein-Rosen bridge connecting the two asymptotic boundaries. That is, any finite stretch of this bridge rapidly shrinks to an exponentially small size. This is universal behavior that we will see both analytically and numerically. Following this rapid collapse, the solution approaches a spacelike Kasner singularity. These regimes are illustrated in Fig. \ref{fig:summary}. With a larger source there is a smoother transition between the RN-AdS and Kasner epochs.

Although Schwarzschild-AdS also has a Kasner singularity, we will see that the singularities that arise from deformations of RN-AdS have Kasner exponents that are bounded away from that of Schwarzschild-AdS. For small deformations,  the Kasner singularity is almost null and `bends up' in the Penrose diagram, while for sufficiently large deformations it can `bend down' like the Schwarzschild-AdS case. Finally, we will also discuss the singularities inside Einstein-Maxwell-dilaton AdS black holes. In these theories, there are analytically known black hole solutions that are free of Cauchy horizons and exhibit spacelike Kasner singularities. They describe the near-horizon geometry of near-extremal black holes, and asymptotically approach a Lifshitz solution. We show that the fixed Kasner exponent of the explicit solutions arises due to an attractor mechanism. Deformations of the Lifshitz regime result in Kasner exponents that depend on the deformation and black hole parameters just like the nondilatonic black holes.

\section{Background and equations}

In the grand canonical ensemble the  dual field theory is held at a chemical potential $\mu$ for some global $U(1)$ symmetry. In the bulk we must correspondingly introduce a Maxwell field $A$ such that $A_t \to \mu$ at the boundary. To deform the boundary theory by a scalar operator $\ocal$ we must introduce a dual scalar field $\phi$ in the bulk.  The leading asymptotic behavior  $\phi_{(0)}$ of the scalar field will be the source for operator. A minimal bulk theory that contains these ingredients is
\be\label{eq:action}
\Lag = R + 6 - \frac{1}{4} F^2 - g^{ab} \pa_a \phi \pa_b \phi - m^2 \phi^2 \,.
\ee
We will consider this theory in $3+1$ bulk dimensions, though as we note below our results hold in higher dimensions also.
We have set the AdS radius and the gravitational coupling to one. The mass $m$ will determine the scaling dimension $\Delta$ of the operator $\ocal$ through the usual formula:
\be
\Delta = \frac{3}{2} + \sqrt{\frac{9}{4} + m^2} \,.
\ee
The Maxwell field strength is $F = dA$.

We wish to find planar charged black hole solutions to the theory (\ref{eq:action}). We will assume the solutions are static and homogeneous, so they can be written in the form
\be\label{eq:metric1}
\mathrm{d}s^2 = \frac{1}{z^2} \left( - f(z) e^{-\chi(z)} \mathrm{d}t^2 + \frac{\mathrm{d}z^2}{f(z)} + \mathrm{d}x^2 + \mathrm{d}y^2  \right) \,,
\ee
The AdS boundary is at $z=0$ and the singularity will be at $z \to \infty$. At a horizon, $f = 0$. The scalar field and scalar potential take the form
\be
\phi = \phi(z) \,, \qquad A = \Phi(z)\,\mathrm{d}t  \,.
\ee
In this gauge regularity requires $\Phi = 0$ at a horizon.
As we will be especially interested in the behavior of the solution behind the horizon, we rewrite the metric in ingoing coordinates:
\be\label{eq:metric}
\mathrm{d}s^2 = \frac{1}{z^2} \left( - f(z) e^{-\chi(z)} \mathrm{d}v^2 - 2 e^{-\chi(z)/2} \mathrm{d}v\,\mathrm{d}z + \mathrm{d}x^2 + \mathrm{d}y^2  \right) \,,
\ee

 The radial functions should obey the following leading asymptotic behavior at the AdS boundary as $z \to 0$
\be\label{eq:boundary}
f \to 1 \,, \qquad \chi \to 0 \,, \qquad \Phi \to \mu \,, \qquad \phi \to \phi_{(0)} z^{3-\Delta} \,.
\ee
This behavior fixes the normalization of time on the boundary as well as the chemical potential $\mu$ and source $\phi_{(0)}$ for the dual operator $\ocal$. Because there is no charged matter in the bulk, it will be convenient to introduce the boundary charge density
\be\label{eq:rho}
\rho = - \lim_{z \to 0} \Phi' \,.
\ee

The bulk equations of motion with the above ansatz are written down as follows. First, the Maxwell equation can be integrated once to give
\be\label{eq:E}
\Phi' = - \rho \, e^{-\chi/2} \,.
\ee
Here $\rho$ is a constant, the boundary charge density (\ref{eq:rho}). The remaining minimal set of
equations of motion can be taken to be
\begin{subequations}
\label{eq:alleqs}
\begin{align}
z^2 e^{\chi/2}\left(e^{-\chi/2} z^{-2} f \phi'\right)' & = \frac{m^2}{z^2} \phi \,, \label{eq:scalar} \\
4 z^4 (z^{-3} f)' & = -12 + 2 m^2 \phi^2  + 2 z^2 f (\phi')^2 + z^4 e^{\chi} (\Phi')^2 \,, \label{eq:fp} \\
\chi' &= z (\phi')^2  \label{eq:chi} \,.
\end{align}
\end{subequations}
Using (\ref{eq:E}) and (\ref{eq:chi}) one can eliminate $\Phi$ and $\chi$ from the equations of motion. The substantive problem is therefore to solve (\ref{eq:scalar}) and (\ref{eq:fp}) for $f$ and $\phi$. We can then immediately obtain $\Phi$ and $\chi$.

\section{Horizons}
\label{sec:noinner}

Solutions to the equations of motion with the asymptotics (\ref{eq:boundary}) will typically have a horizon at $z_{\mathcal{H}}$, with $f(z_{\mathcal{H}}) = 0$. The temperature of the dual quantum field theory is
\be
T = \frac{1}{4 \pi} |f'(z_{\mathcal{H}})| e^{-\chi(z_{\mathcal{H}})/2} \,.
\ee
The infalling coordinates (\ref{eq:metric}) continue across the horizon. Our main interest is the interior geometry that is found beyond the horizon.

In the absence of a scalar field, with $\phi = 0$ everywhere, the solution is the Reissner-Nordstr\"om-AdS spacetime, with $\chi = 0$ and
\be
f_\text{RN}(z) = 1 + \frac{\rho^2 z^4}{4} - \left(\frac{z}{z_{\mathcal{H}}}\right)^3 \left(1 + \frac{\rho^2 z_{\mathcal{H}}^4}{4} \right) \,.
\label{eq:fRNADS}
\ee
In addition to the horizon at $z = z_{\mathcal{H}}$, there is  an inner horizon at $z = z_{\mathcal{I}}$ with
\be
\left(\frac{z_{\mathcal{I}}}{z_{\mathcal{H}}}\right)^2 + \frac{z_{\mathcal{I}}}{z_{\mathcal{H}}} + 1 = \frac{\rho^2 z_{\mathcal{H}}^4}{4} \left(\frac{z_{\mathcal{I}}}{z_{\mathcal{H}}}\right)^3 \,.
\ee
This inner horizon is well known to be a Cauchy horizon, leading to the breakdown of predictability in the black hole interior. At high temperatures $\rho^2 z_{\mathcal{H}}^4 \to 0$ and in this limit the inner horizon is at $z_{\mathcal{I}} \approx 4 z_{\mathcal{H}}/(\rho^2 z_{\mathcal{H}}^4) \to \infty$, although the proper time between the horizons does not become large.  At low temperatures $z_{\mathcal{I}} \to z_{\mathcal{H}}$ as the black hole becomes extremal.

We now discuss the effect of a nonzero scalar field on the inner horizon. For the theory with action (\ref{eq:action}) this depends on the sign of the mass squared, which also corresponds to whether the operator is relevant or irrelevant. With a more general potential for the scalar field, however, there need be no connection between relevance or irrelevance near the AdS boundary and the sign of the potential in the black hole interior.

\subsection{Relevant deformations remove Cauchy horizons}\label{sec:proofnoC}

The black hole interior is dramatically changed by a nonzero $\phi$. For $m^2 \leq 0$, which corresponds to relevant operators with $\Delta \leq 3$ in our theory, we can prove that there is no inner horizon as follows. Suppose that there were two horizons at $z_{\mathcal{H}}$ and $z_{\mathcal{I}}$. From  Eq. (\ref{eq:scalar}):
\begin{align}
0 = & \int_{z_{\mathcal{H}}}^{z_{\mathcal{I}}} \left(\frac{f e^{-\chi/2}\phi \phi^\prime}{z^2}\right)^\prime \mathrm{d}z = \int_{z_{\mathcal{H}}}^{z_{\mathcal{I}}} \frac{e^{-\chi/2}}{z^4}  \left[m^2 \phi^2 + z^2 f (\phi^\prime)^2 \right]\,\mathrm{d}z
 \,.
\end{align}
In the first equality we have used the fact that $f(z_{\mathcal{H}}) = f(z_{\mathcal{I}}) = 0$. In the final expression note that between the two horizons $f < 0$. If $m^2 \leq 0$, the integrand in the final expression is therefore non-positive over the range of integration. Thus, the only way there can be two horizons is if $\phi = 0$ identically. The scalar field necessarily removes the inner horizon. For more general scalar potentials $V(\phi)$, the above argument still applies provided $\phi V'(\phi) < 0$.

\subsection{Irrelevant deformations can have fine-tuned Cauchy horizons}
\label{sec:rel}

For certain irrelevant deformations, we will see that  inner horizons can exist at one specific temperature. Irrelevant operators are dual to bulk fields with $m^2 > 0$. These grow large towards the AdS boundary, and so cannot be consistently included as sources. Instead they will induce a renormalization group flow up towards a different UV completion. Our analysis will only depend on the scalar field profile in between the black hole horizon and the Cauchy horizon, and is therefore independent of the UV completion. We will do this in two steps: first we analyse the linear problem, and then we bootstrap the problem non-linearly.

For the linearized problem we look at the scalar field 
on the Reissner-Nordstr\"om background. This amounts to 
Eq.~(\ref{eq:scalar}) with $\chi=0$ and $f = f_\text{RN}$ as in (\ref{eq:fRNADS}):
\begin{equation}
z^4\left(z^{-2} f_\text{RN} \phi^\prime\right)^\prime = m^2\phi \,,
\label{eq:simplescalar}
\end{equation}
We wish to solve (\ref{eq:simplescalar}) for $z\in(z_{\mathcal{H}},z_{\mathcal{I}})$ --- where here 
$z_{\mathcal{H}}$ and $z_{\mathcal{I}}$ are the outer and inner horizons of RN-AdS --- with the regularity conditions that
\be
\phi^\prime(z_{\mathcal{H}})=\frac{m^2}{z_{\mathcal{H}}^2f_\text{RN}^\prime(z_{\mathcal{H}})}\phi(z_{\mathcal{H}})\,,\quad \text{and} \quad \phi^\prime(z_{\mathcal{I}})=\frac{m^2}{z_{\mathcal{I}}^2f_\text{RN}^\prime(z_{\mathcal{I}})}\phi(z_{\mathcal{I}})\,.
\ee
These boundary conditions together with (\ref{eq:simplescalar}) define an eigenvalue problem for $m^2$. Because $f_{\text{RN}}<0$ between the two horizons, it is clear from (\ref{eq:simplescalar}) that there are no eigenvalues with $m^2 < 0$, consistent with our result in the previous section. Perhaps unsurprisingly, we will find an infinite tower of positive eigenvalues of $m^2$. The eigenvalues can be written as a function of
\be
\xi \equiv \frac{z_{\mathcal{I}}}{z_{\mathcal{H}}} \geq 1 \,.
\ee
 Given a UV completion that restores an asymptotically AdS region, for instance due to a more complicated scalar potential than just $m^2 \phi^2$, the ratio $\xi$ has the same information as the dimensionless boundary quantity $T/\mu$. (The asymptotic region is necessary to fix the normalization of the time coordinate.) At extremality, $\xi = 1$.

\begin{figure}[h]
\centering
\includegraphics[width=0.6\linewidth]{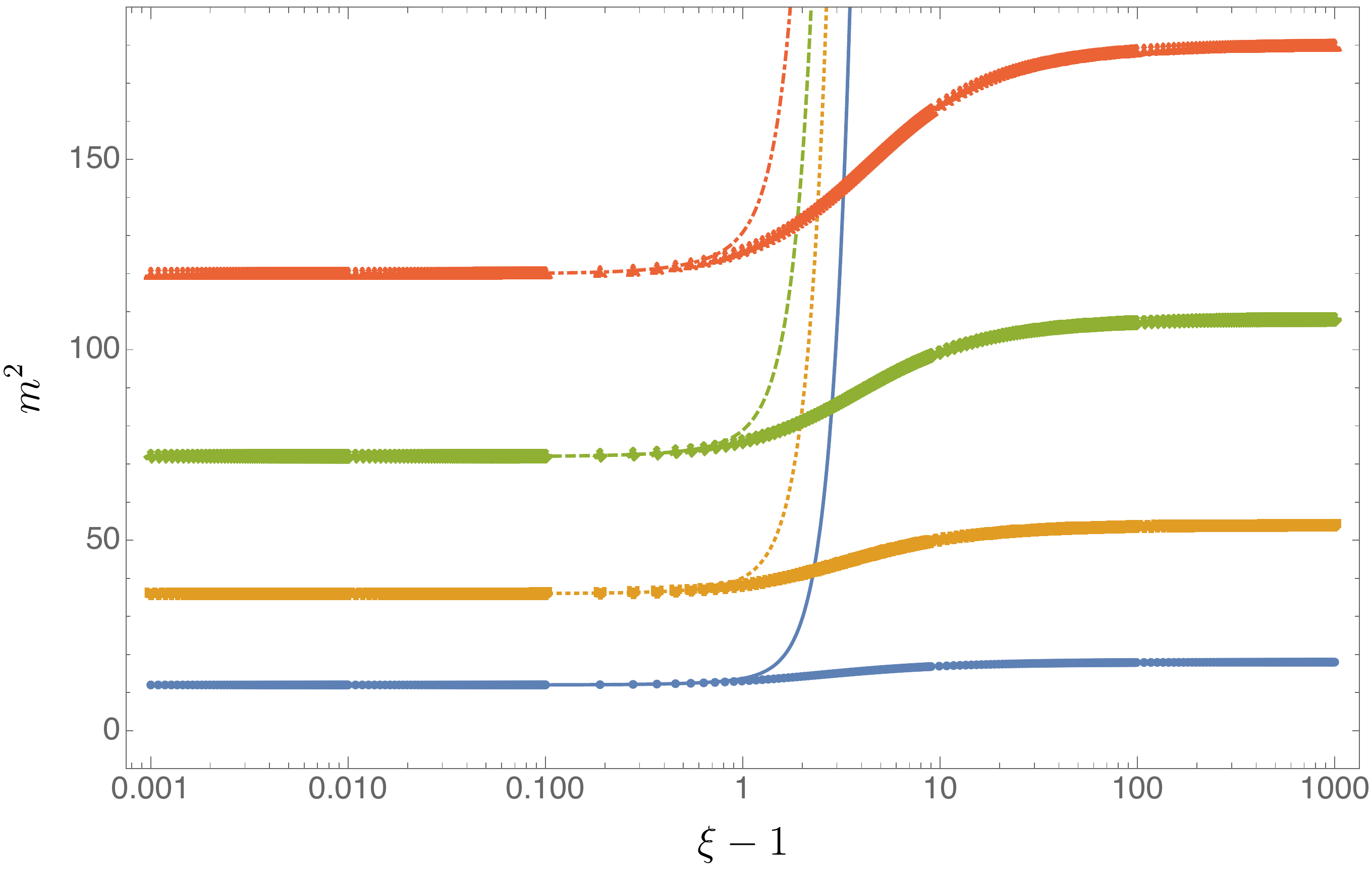}
\caption{The values of mass squared allowing for a regular scalar field between the inner and outer horizons of RN-AdS, as a function of the near-extremality parameter $(\xi-1)$. The lowest few solutions of the infinite tower are shown. The solid, dashed, dotted and dotted-dashed lines correspond to the perturbative result given by (\ref{eq:m2ext}) with $\ell=1,2,3,4$, respectively. The disks, squares, diamonds and triangles are the corresponding exact numerical data.}
\label{fig:m2}
\end{figure}

The linearized eigenvalue problem can be readily solved via the numerical methods detailed in \cite{Dias:2015nua}. Alternatively, we can perturbatively solve (\ref{eq:simplescalar}) around extremality, using the methods of \cite{Dias:2018ufh}. As expected, we find an infinite tower of modes, which we label by an integer $\ell\geq1$. For these masses, a regular scalar field configuration exists between the inner and outer horizon. We shall just quote here the result for $m^2$ to quartic order in $(\xi-1)$ for generic values of $\ell$. Once the dust settles, we find:
\begin{multline}
    m^2 = 6 \lambda _{\ell } \Bigg[1+\frac{5 \lambda _{\ell }+2}{12 (2 \ell -1) (2 \ell +3)}(\xi-1)^2-\frac{5 \lambda _{\ell }+2}{12 (2 \ell -1) (2\ell +3)}(\xi-1)^3+
   \\
   \frac{18460 \lambda _{\ell }^4-82565 \lambda _{\ell }^3+60864 \lambda _{\ell }^2+13608 \lambda _{\ell }-11880}{864 (2 \ell -3) (2 \ell +5) (2 \ell-1)^3 (2 \ell +3)^3} (\xi-1)^4
    \Bigg]+\mathcal{O}\left[(\xi-1)^5\right]\,,
    \label{eq:m2ext}
\end{multline}
where $\lambda_{\ell}=\ell(\ell+1)$. In Fig.~\ref{fig:m2} we show the numerically determined values of $m^2$ as a function of $(\xi-1)$. The numerical and perturbative results agree for $\xi \sim 1$.

The large $\xi$ behavior shown in Fig.~\ref{fig:m2} can also be understood analytically. In the strict $\xi \to \infty$ limit, the RN-AdS background becomes Schwarzschild-AdS.   Generically, linear massive scalar fields in this spacetime diverge logarithmically near the singularity. (This leads to a change in the Kasner exponents in the full nonlinear solutions, as discussed in \cite{Frenkel:2020ysx}.) The analog of demanding that the Cauchy horizon remain smooth, is to demand  that the scalar field vanish at the singularity. If one imposes this (and regularity at the event horizon), one again obtains an eigenvalue equation for $m^2$ with  eigenvalues:
\be
{m^2  = 9 \,\lambda_{\ell}+\mathcal{O}(\xi^{-1})} \,.
\ee
The large $\xi$ results in Fig.~\ref{fig:m2} indeed asymptote to these values. One can go a bit further and compute the corrections in $\xi^{-1}$ using standard perturbation theory. These turn out to be given by
\be
m^2= 9 \,\lambda_{\ell}\left[1-\frac{2 \ell+1}{\xi}\frac{\Gamma \left(\frac{5}{3}\right) \Gamma \left(\ell+\frac{1}{3}\right)}{\Gamma \left(\frac{1}{3}\right)
   \Gamma \left(\ell+\frac{5}{3}\right)}{}_4F_3\left(1-\ell,2+\ell,\frac{2}{3},\frac{5}{3};\frac{2}{3}-\ell,\frac{5}{3}+\ell,2;1\right)\right]+\mathcal{O}(\xi^{-2})\,,
\ee
where $_4F_3\left(a,b,c,d;e,f,g;z\right)$ is a generalised hypergeometric function.

Since $m^2$ is a parameter in the bulk action, it is probably more physical to turn Fig.~\ref{fig:m2} around and interpret it as saying that for certain given $m^2$, there can be  one value of $T/\mu$ for which the inner horizon is not destroyed (at the linearized level).

We now establish that these linearized solutions extend without obstruction to nonlinear solutions with a smooth Cauchy horizon. As noted below (\ref{eq:alleqs}) the equations to be solved are a first order equation for $f$ and a second order equation for $\phi$. There are correspondingly three constants of integration. We can take these to be $\{\xi,\phi_{\mathcal{H}},\phi_{\mathcal{I}}\}$. Here $\phi_{\mathcal{H}} = \phi(z_{\mathcal{H}})$ and $\phi_{\mathcal{I}} = \phi(z_{\mathcal{I}})$. These equations in addition depend on the parameters $\rho z_{\mathcal{H}}^2$ and $m^2$. A solution can therefore be specified by the five parameters $\{m^2,\rho z_{\mathcal{H}}^2,\xi,\phi_{\mathcal{H}},\phi_{\mathcal{I}}\}$. Suppose that we take a solution that is regular at the outer horizon and integrate in, and we take a solution that is regular at the inner horizon and integrate out. These will combine into a solution that is regular everywhere between the horizons if $\{\phi, \phi^\prime,f\}$ match at some intermediate point. With five paramaters and three constraints we expect to find a two-parameter family of solutions with a smooth Cauchy horizon. These can be labelled \emph{e.g.} by $\{\xi, \phi_{\mathcal{H}}\}$. As  $\phi_{\mathcal{H}} \to 0$, $m^2$ should match the values obtained previously from a linearized analysis in Fig.~\ref{fig:m2}.

\begin{figure}[h]
\centering
\includegraphics[width=\linewidth]{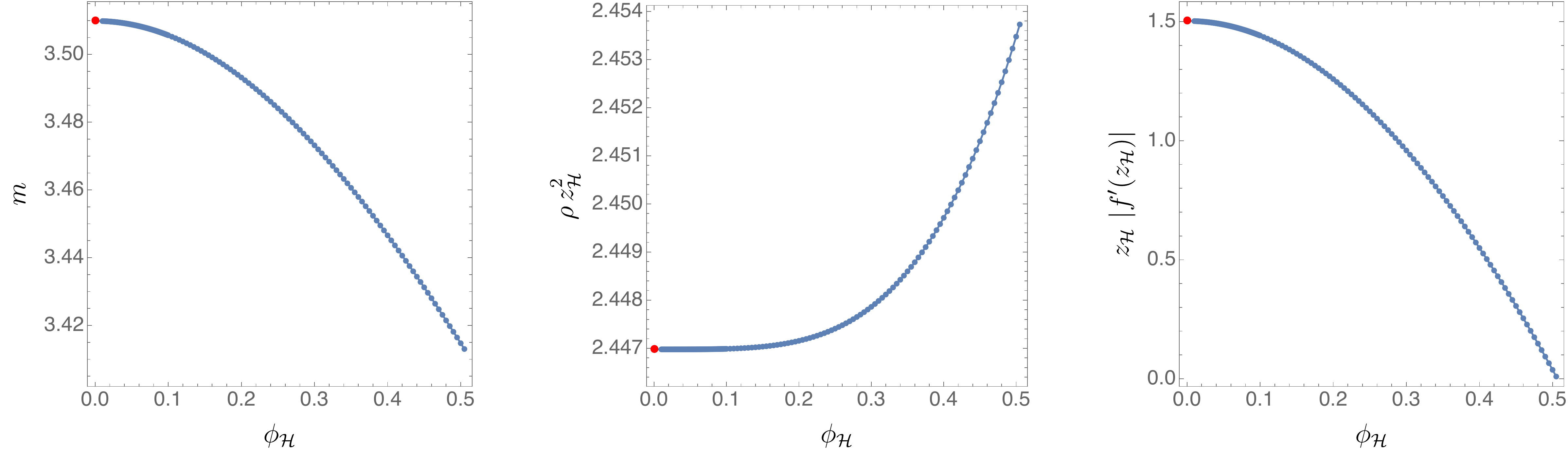}
\caption{Nonlinearly smooth Cauchy horizons as a function of the scalar field at $z_{\mathcal H}$. \textbf{Left:} mass $m$ of the scalar field, \textbf{middle:} charge density $\rho$, \textbf{right:} $z_{\mathcal{H}}\left|f^\prime(z_{\mathcal{H}})\right|$. All plots were generated while keeping $\xi=1.448$. The red disks were obtained by solving the linear problem.}
\label{fig:example}
\end{figure}

We have scanned a large portion of parameter space, and found the above counting picture to be correct. In Fig.~\ref{fig:example} we show an example at fixed $\xi=1.448$. This leaves a one parameter family of solutions that extend the linearized solutions to nonzero $\phi_{\mathcal{H}}$. On the left panel we plot the mass, on the middle panel the charge density and on the right panel $z_{\mathcal{H}}\left|f^\prime(z_{\mathcal{H}})\right|$, with the latter quantity being proportional to the black hole temperature. The final plot suggests that at fixed $\xi$ we can always find a large enough value of $\phi_{\mathcal{H}}$ where we reach extremality (and $f$ acquires a double zero).
Furthermore, we have checked that the extremal limit appears non-singular, in the sense that $R_{abcd}R^{abcd}$ does not appear to blow up when $z_{\mathcal{H}}\left|f^\prime(z_{\mathcal{H}})\right|\to0$, nor any other curvature invariant. In addition, we searched for tidal force singularities, and found none. We have chosen many other values of $\xi$, and the overall behaviour appears similar.

\section{Collapse of the Einstein-Rosen bridge}\label{sec:collapse}

When the inner horizon is absent, the black hole interior ends at a spacelike singularity  as $z \to \infty$. We describe the asymptotic near-singularity behavior in the following section. In this section we describe a crossover that occurs at the location $z_{\mathcal{I}}$ of the would-be horizon. The crossover is most dramatic when the scalar field is small, and in this limit can be obtained analytically. While the scalar field is small, the spacetime dynamics is highly nonlinear in this regime. We will see that it corresponds to a collapse of the Einstein-Rosen bridge between the two asymptotic boundaries.

The collapse occurs over an extremely short range in the $z$ coordinate, so it is consistent to simply set $z \to z_{\mathcal{I}}$ in the equations of motion (\ref{eq:scalar}) -- (\ref{eq:chi}). We can think of the variables $f,\chi,\phi$ as functions of $\delta z = z - z_{\mathcal{I}}$. Furthermore, at these values of $z$ it can be verified numerically (or, a posteriori on the solution below) that the mass of the scalar field becomes negligible in (\ref{eq:scalar}) and (\ref{eq:fp}). With these approximations 
the equations become
\be
\left(e^{-\chi/2} f \phi'\right)' = 0 \,, \qquad 
4 z_{\mathcal{I}} f' = 2 z_{\mathcal{I}}^2 f (\phi')^2 + z_{\mathcal{I}}^4 \r^2 - 12 \,, \qquad
\chi' = z_{\mathcal{I}} (\phi')^2 \,. \label{eq:th}
\ee
The general solution to these equations can be found, starting by integrating the first equation and writing  $\phi' = - c_1 (z_{\mathcal{I}}^4 \rho^2/2 - 6)^{1/2} e^{\chi/2}/f$. Here $c_1$ is a constant and the normalization is for future convenience.
The solution is most nicely expressed in terms of the metric component $g_{tt} = -f e^{-\chi}/z_{\mathcal{I}}^2$.
This is found to obey
\be
\frac{g_{tt}''}{g_{tt}'} = \frac{c_1^2 g_{tt}'}{g_{tt} \left(c_1^2 + g_{tt} \right)} \,.
\ee
The general solution to this equation takes the form (recall $g_{tt} > 0$ in the black hole interior) 
\be\label{eq:gtt}
c_1^2 \log (g_{tt}) + g_{tt} = -\frac{z_{\mathcal{I}}}{2} \, c_2^2(\delta z + c_3) \,.
\ee
Here $c_2$ and $c_3$ are additional constants of integration (again normalized for convenience).
And then, in addition to $f = - e^{\chi} g_{tt} z_{\mathcal{I}}^2$, one finds that
\be\label{eq:phichi}
\phi = - \frac{2 c_1}{z_{\mathcal{I}} c_2} \log \left(c_4 \, g_{tt}\right) \,, \qquad e^{-\chi} = \frac{2 z_{\mathcal{I}}^4}{c_1^2 (z_{\mathcal{I}}^4 \rho^2-12)} (\phi')^2 g_{tt}^2 \,.
\ee
The scalar field exhibits the expected logarithmic growth as $g_{tt}$ becomes small close to the would-be inner horizon. The special cases discussed in the previous section where the inner horizon survives will have $c_1 = 0$.

\begin{figure}[h]
\centering
\includegraphics[width=0.95\linewidth]{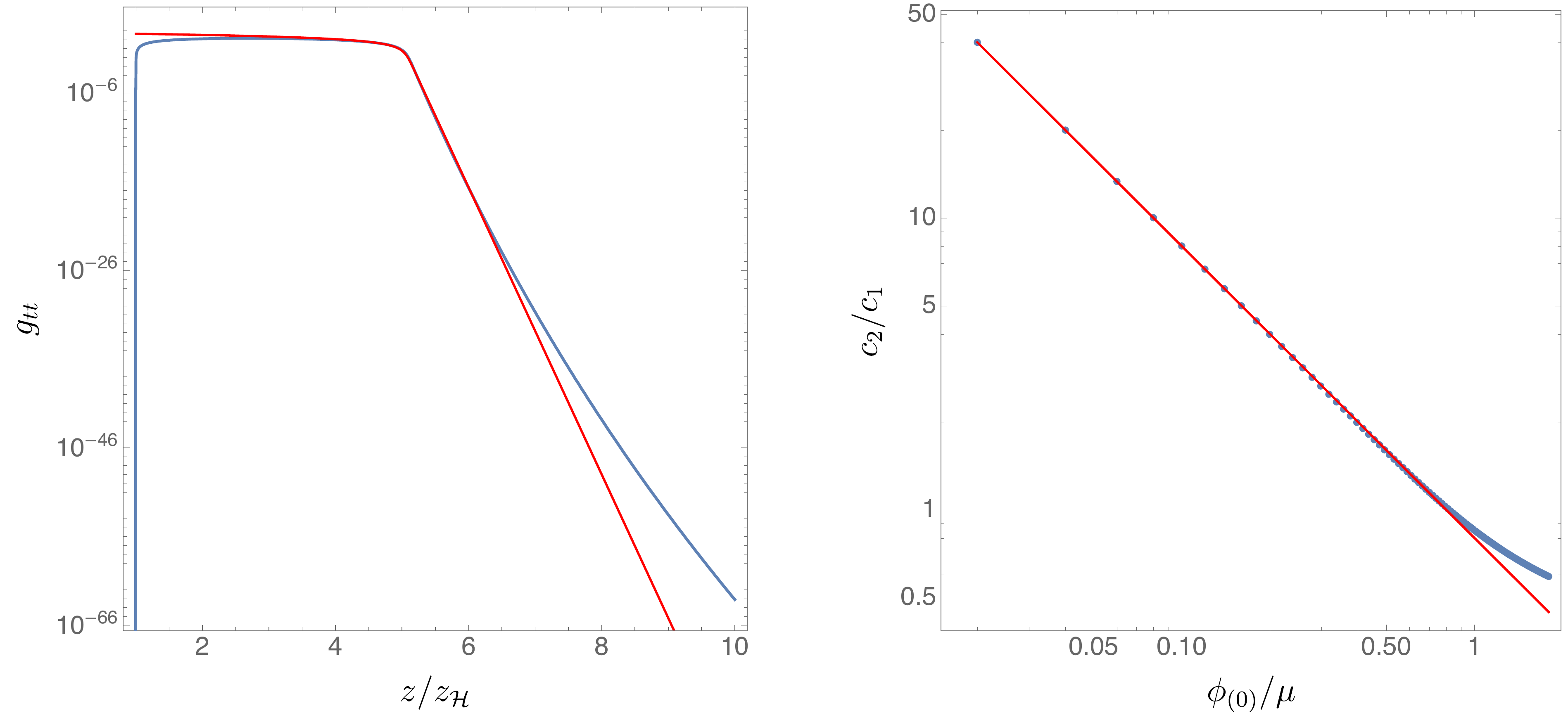}
\caption{Collapse of the Einstein-Rosen bridge. {\bf Left:} abrupt crossover of $g_{tt}$ at the would-be inner horizon. Blue line is from a numerical solution of the equations of motion and the red line is a fit to the analytic crossover form (\ref{eq:gtt}). {\bf Right:} fit parameter $c_2/c_1$ as a function of the boundary value of the scalar deformation. As the deformation becomes small, $c_2/c_1$ becomes large and the crossover more dramatic. Blue dots are numerical data points and the red line shows $c_2/c_1 \approx 0.8 \mu/\phi_{(0)}$. All data shown for temperature $T/\mu \approx 0.2188$ and a scalar field with $m^2 = -2$. Left plot has $\phi_{(0)}/\mu \approx 0.2193$, corresponding to $c_2/c_1 \approx 3.89$.
}
\label{fig:collapse}
\end{figure}

The first equation in (\ref{eq:phichi}) suggests that $c_2/c_1$ will become large when the boundary source for $\phi$ is small. This is because the argument of the logarithm in (\ref{eq:phichi}) is order one at the end of the crossover region, and in the limit of a small scalar field, the scalar can be integrated from the crossover region to the asymptotic boundary as a linear equation. We verify from numerics in Fig. \ref{fig:collapse} that indeed $c_2/c_1 \sim 1/\phi_{(0)}$ as the source $\phi_{(0)} \to 0$.  Thus even while $\delta z$ is small, $(c_2/c_1)^2 \delta z$ in (\ref{eq:gtt}) can be very large. This allows the metric to undergo a big change with the coordinate $z$ hardly changing. This fact is, a posteriori, what has allowed us to only solve the equations in the vicinity of $z_{\mathcal{I}}$. A large $c_2/c_1$ in (\ref{eq:gtt}) leads to an extremely fast crossover in behavior (setting the shift $c_3 = 0$ here for clarity):
\begin{align}
\delta z <0 \quad & \to \quad \delta z > 0 \,, \nonumber \\
g_{tt} = \frac{z_{\mathcal{I}} c_2^2}{2} |\delta z| \quad & \to \quad g_{tt} = e^{-(c_2^2 z_{\mathcal{I}}/2 c_1^2) \, \delta z} \,,  \label{eq:gttcross}\\
\phi' = \frac{c_1}{c_2} \frac{1}{|\delta z|} \quad & \to \quad \phi' = \frac{c_2}{c_1} \,.
\end{align}
Here we see that a linear vanishing of $g_{tt}$ towards the would-be inner horizon is replaced by a rapid collapse to an exponentially small value, while the divergence in the scalar field derivative towards the horizon is cut off at a large value. This behavior is verified by comparison with numerical solutions to the equations of motion, illustrated in Fig. \ref{fig:collapse}. The inversion in the value of the scalar derivative reveals the nonlinear nature of this transition.

In the black hole interior, $g_{tt}$ sets the measure for the spatial $t$ coordinate that runs along the wormhole connecting the two exteriors of the black hole. This is the Einstein-Rosen bridge. The rapid decrease in $g_{tt}$ that we have just described can therefore be thought of as a collapse of the Einstein-Rosen bridge for a fixed coordinate separation $\Delta t$. The collapse to an exponentially small $g_{tt}$ happens over a short proper time $\propto c_1^3/c_2^3$.

\section{Kasner singularity}\label{sec:Kasner}

After the collapse of the Einstein-Rosen bridge, the spacetime enters an asymptotic regime that tends towards a Kasner singularity. Recall that the Kasner solution is a homogeneous, anisotropic cosmology with power law behavior near the singularity. When the Maxwell flux terms are subleading, the asymptotic solution is given by 
\cite{Kasner, Belinski:1973zz}
\be\label{eq:kasner}
\mathrm{d}s^2 = - \mathrm{d}\tau^2 + c_t \tau^{2 p_t} \mathrm{d}t^2 + c_x \tau^{2 p_x} \left(\mathrm{d}x^2 + \mathrm{d}y^2 \right) \,, \qquad \phi = - \sqrt{2} p_\phi \log \tau \,.
\ee
Here $c_t$ and $c_x$ are constants.
The Kasner exponents obey $p_t + 2 p_x = 1$ and $p_\phi^2 + p_t^2 + 2 p_x^2 = 1$.

\begin{figure}[ht!]
    \vskip 0.1in
    \centering
    \includegraphics[width=0.7\textwidth]{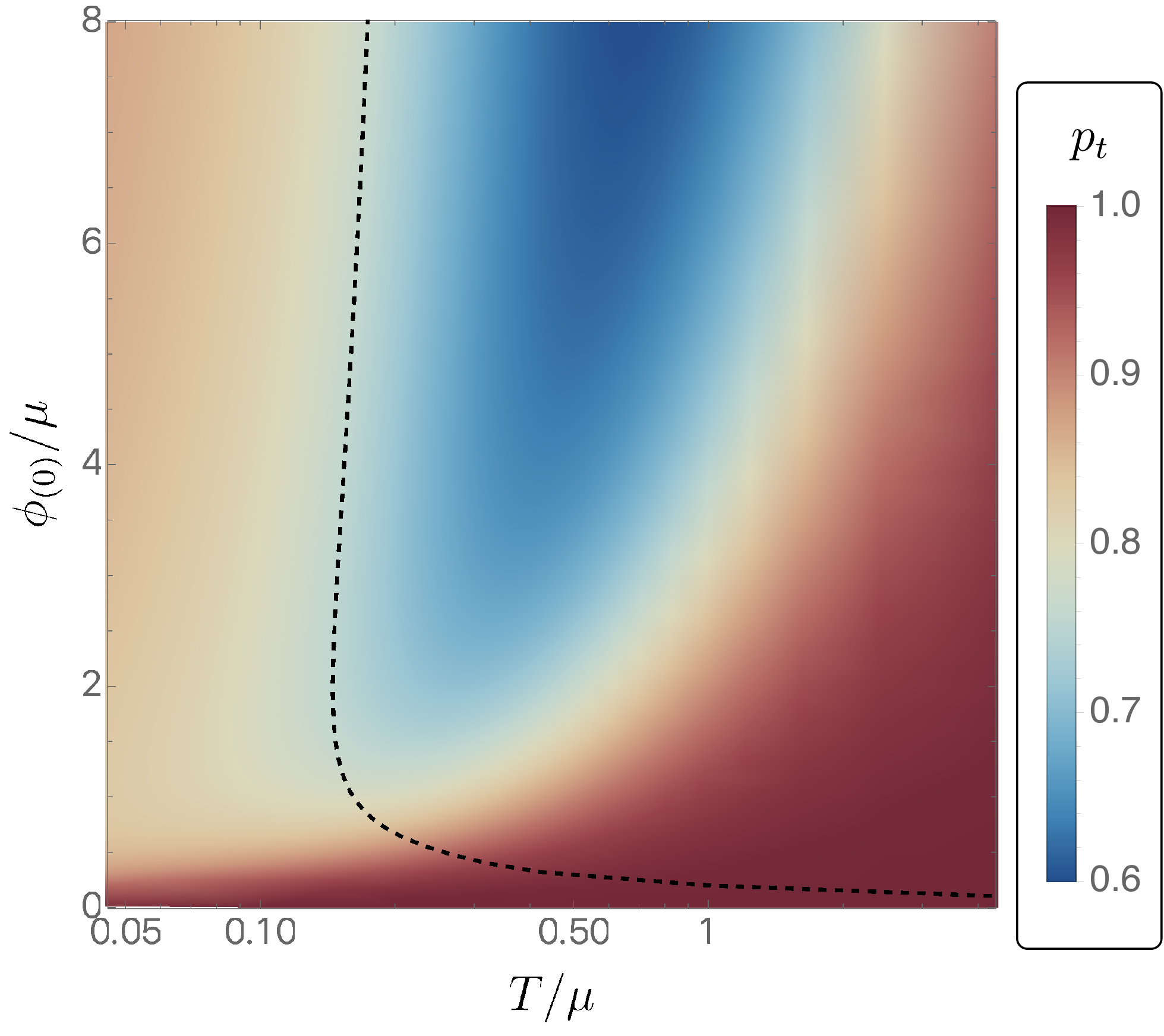}
    \caption{Near-singularity Kasner exponent $p_t$ as a function of the boundary temperature $T/\mu$ and strength of the deformation $\phi_{(0)}/\mu$. The scalar field is taken to have $m^2 = -2$. The dashed line shows the transition from the singularity bending up to bending down in the Penrose diagram, discussed in \S\ref{sec:Penrose}.}
    \label{fig:phaseDiag}
    \vskip -0.1in
\end{figure}

The near-singularity behavior is similar to that of the neutral black holes studied in \cite{Frenkel:2020ysx}. We find that as $z \to \infty$, the solutions
 take the form
\begin{align}
f  = - f_o z^{3 + \a^2} \,, \quad \phi  = \a \sqrt{2} \log z  \,, \quad \chi = 2 \a^2 \log z + \chi_o  \,, \quad
\Phi' = - \rho e^{-\chi_o/2} z^{-\a^2} \,, \label{eq:kas}
\end{align}
with $\alpha > 1$. This restriction on $\alpha$ ensures that the Maxwell flux terms are always unimportant asymptotically.
It is easy to see that the metric and scalar are indeed of the Kasner form (\ref{eq:kasner}) with 
\be\label{eq:pt}
p_t = \frac{\a^2 - 1}{3 + \a^2} \,, \qquad 1 \geq p_t > 0 \,.
\ee
The lower bound on $p_t$ (following from the bound on $\a$) excludes the Schwarzschild near-singularity behavior which has $p_t = -1/3$. Fig. \ref{fig:phaseDiag} shows $p_t$ as a function of the boundary value of the scalar
field and the temperature, for the choice of mass $m^2 = -2$.

Fig. \ref{fig:phaseDiag} shows that,
consistent with our proof of no inner horizon for $m^2 < 0$ in \S \ref{sec:noinner}, the entire $\phi_{(0)}>0$ and $T>0$ phase diagram flows to a spacelike Kasner singularity. We now describe the limits $\phi_{(0)} \to 0$ and $T \to 0$.

\begin{figure}[ht!]
    \vskip 0.1in
    \centering
    \includegraphics[width=0.465\textwidth]{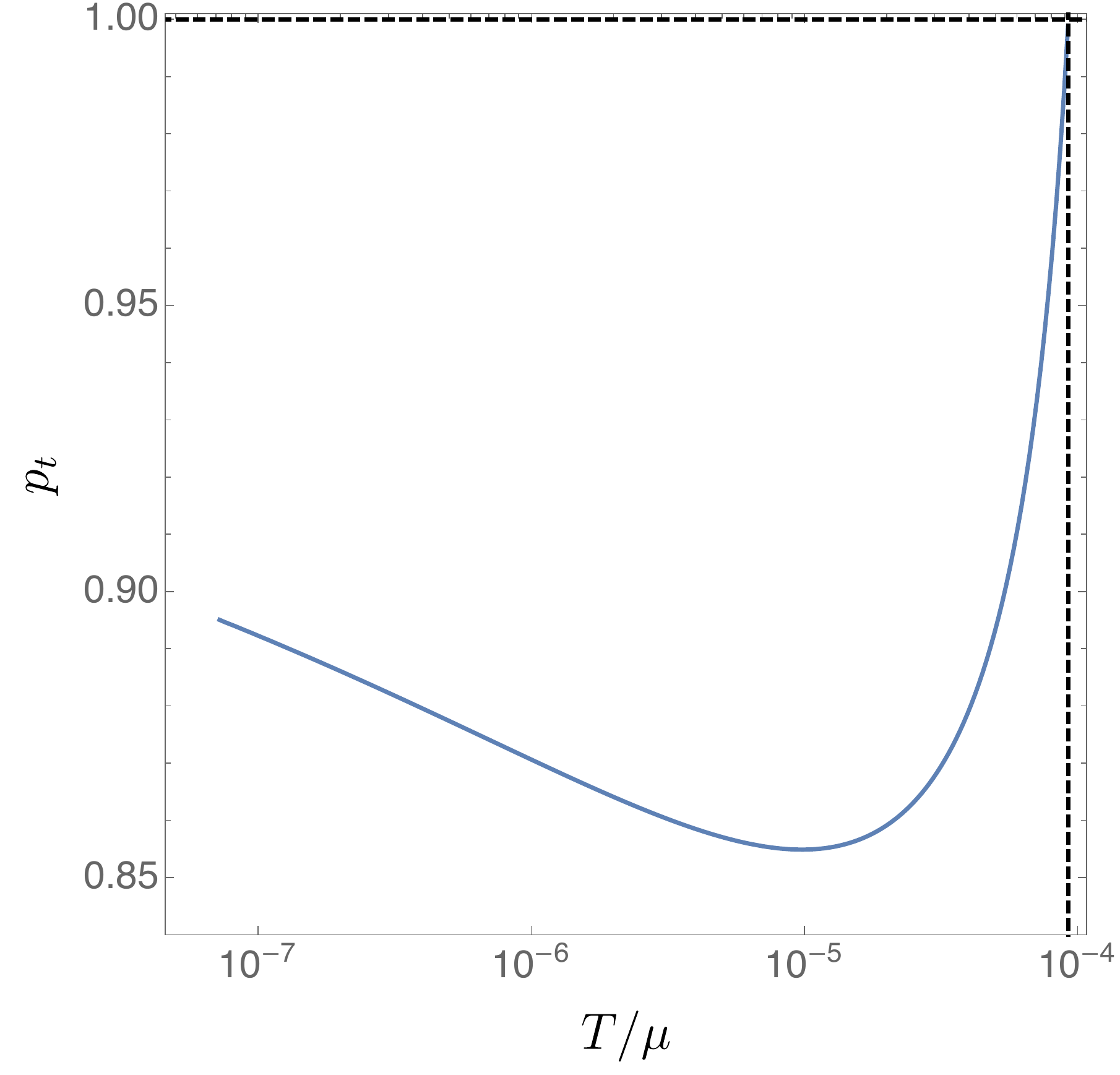}
    \hspace{0.5cm}
    \includegraphics[width=0.45\textwidth]{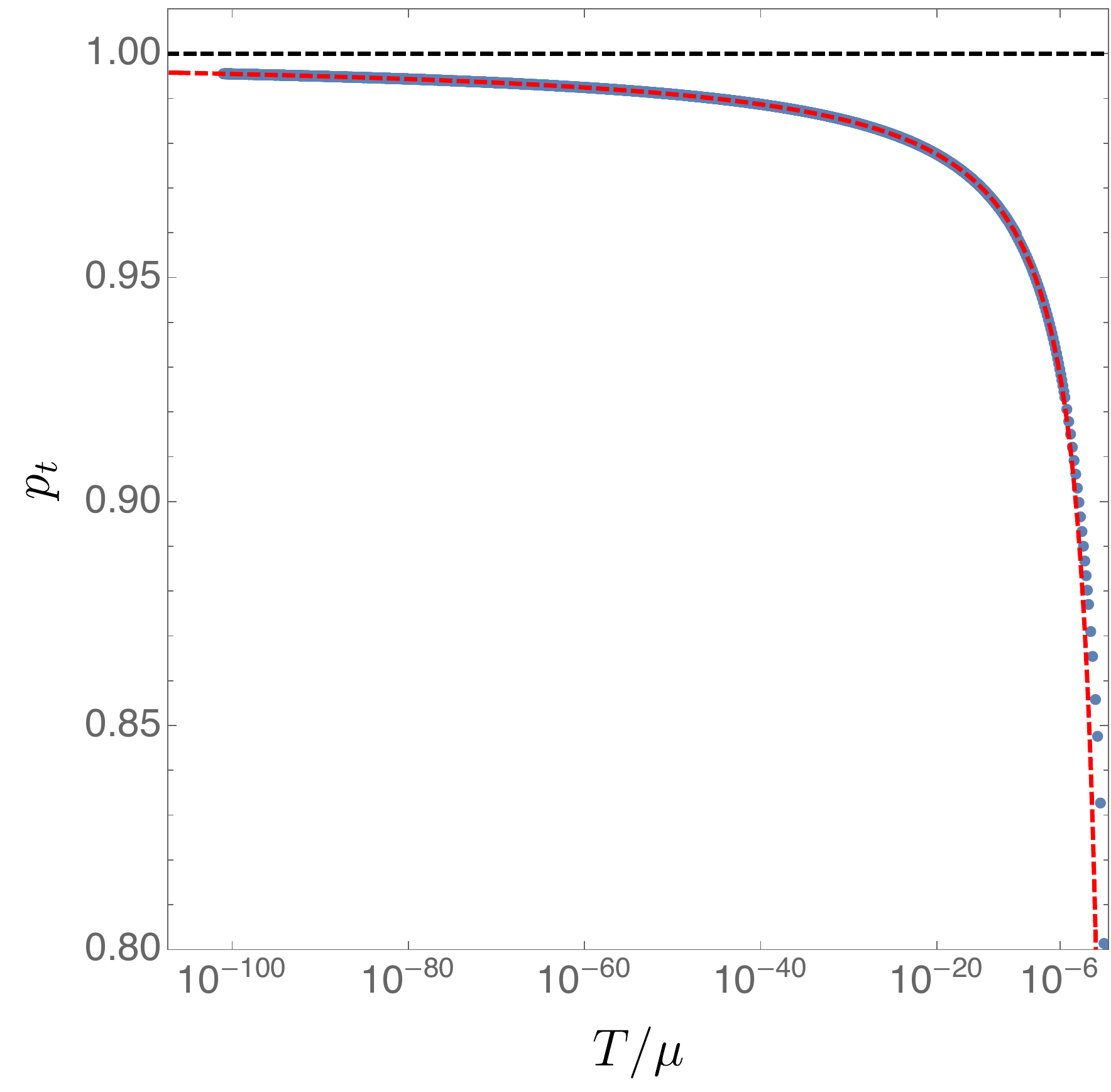}
    \caption{{\bf Left}: Near-singularity Kasner exponent $p_t$ at $\phi_{(0)}=0$ below the temperature $T_c = 9.26 \times 10^{-5} \mu$ (vertical dashed line) at which the scalar spontaneously condenses. For $T>T_c$, $p_t =1$. {\bf Right}: Near-singularity Kasner exponent $p_t$ in a linear$-\log$ scale as a function of $T/\mu$, at constant $\phi_{(0)}/\mu = 1$. The red dashed line shows a one-parameter fit to (\ref{eq:pt}) with $\alpha = a_0\,[- \log(T/\mu)]^{1/2}$ in the range $T/\mu \in(10^{-100},10^{-6})$. For this run, we obtained $a_0\simeq 1.94876$.}
    \label{fig:pt_Spont_nonSpont}
    \vskip -0.1in
\end{figure}

As the scalar field is turned off at fixed temperature, the Kasner exponent $p_t \to 1$. This is different to the case of neutral black holes, where $p_t \to -1/3$ as the deformation is turned off \cite{Frenkel:2020ysx}. The difference is easily understood: In the neutral case the solution reverts to the Schwarzschild singularity, while in the charged case the Kasner singularity reverts to the regular inner horizon (which has $p_t = 1$ in Kasner coordinates). An exception to this statement arises at very low temperatures. At sufficiently low temperatures, neutral scalar fields can spontaneously condense in the Reissner-Nordstr\"om-AdS background \cite{Hartnoll:2008kx}. Below the critical temperature $T_c$, this leads to a Kasner singularity with $p_t < 1$ even in the absence of a source, $\phi_{(0)} = 0$, as shown in Fig. \ref{fig:pt_Spont_nonSpont}. Each spontaneous solution will extend to a family of solutions with nonzero source. For small values of the source, these solutions will compete with the solutions that continue to the trivial solution at $\phi_{(0)} = 0$.

As $T/\mu \to 0$ for fixed deformation strength $\phi_{(0)}/\mu$, again the Kasner exponent $p_t \to 1$. This is not completely clear from Fig. \ref{fig:phaseDiag} but is seen clearly in Fig. \ref{fig:pt_Spont_nonSpont}. We make two further observations. Firstly, in Fig. \ref{fig:pt_Spont_nonSpont} we see that extending to very low temperatures, the limiting numerical behavior is well fit by $\alpha \propto [- \log(T/\mu)]^{1/2} \to \infty$ in (\ref{eq:pt}). This corresponds to $p_t \to 1$. Secondly, in this limit the outer horizon is verified in numerics to be approaching a singular solution first written down in \cite{Horowitz:2009ij}:
\begin{align}\label{eq:sing}
f = \frac{(m^2)^2}{2} \frac{1}{\rho^2 z^4} + \cdots \,, \qquad
\phi = \pm \frac{\rho z^2}{\sqrt{- 2 m^2}} + \cdots \,.
\end{align}
The series expansion continues in powers of $1/(\r z^2)^2$. In addition, there is a nonperturbative contribution of the form $\delta \phi = A\, \exp\left\{-\frac{\rho\,z^2}{\sqrt{6+m^2}}\right\}$.\footnote{This corrects a statement in \cite{Horowitz:2009ij}.} The parameter $A$ is fixed by the asymptotic source, $\phi_{(0)}$.
In the low temperature limit $\r z_{\mathcal H}^2 \to \infty$. This allows the expansion above in $\rho z^2 \gg 1$ even outside the outer horizon where $z < z_{\mathcal H}$. The
divergence of $\a$ in the low temperature limit is consistent with the scalar field $\phi$ crossing over to the stronger than logarithmic growth of (\ref{eq:sing}).

Finally, we note that in the discrete cases with $m^2 > 0$ where an inner horizon survives, as discussed in \S \ref{sec:rel}, the singularity beyond the inner horizon will be that of the Reissner-Nordstr\"om black hole, with the scalar field becoming unimportant towards the singularity ($\phi \sim 1/z$) and the equations dominated by the flux terms.

\section{Penrose diagrams}\label{sec:Penrose}

Penrose diagrams are convenient ways to picture the global causal structure of a spacetime.  Given a static AdS black hole with a spacelike singularity, one often imagines its Penrose diagram  is a square, with singularities on top and bottom. However, as pointed out in \cite{Fidkowski:2003nf} there is a conformally invariant distinction between spacelike singularities that bend down toward the event horizon and ones that bend up away from the horizon. In Schwarzschild-AdS, the singularity bends down \cite{Fidkowski:2003nf}. For small deformations of the RN-AdS spacetime, the singularity appears close to the would-be inner Cauchy horizon and hence one might expect the singularity to bend upwards in this limit. Let us now discuss this more systematically.

\begin{figure}[ht!]
    \vskip 0.1in
    \centering
    \includegraphics[width=0.5\textwidth]{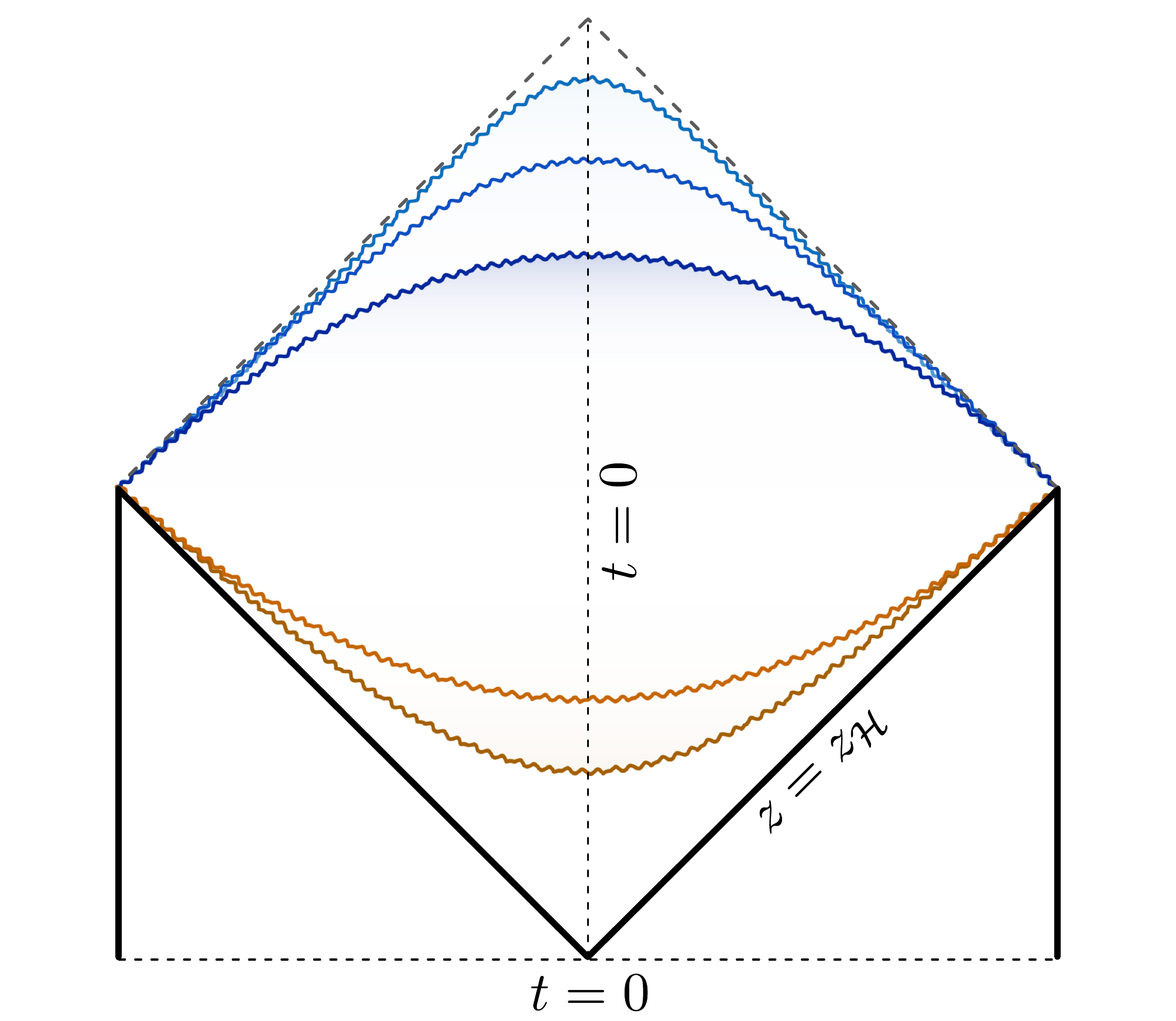}
    \caption{Penrose diagrams for the grand canonical ensemble. At low temperatures or for small deformations, the singularity bulges up and approaches the Cauchy horizon of the Reissner-Nordstr\"{o}m solution, shown with grey dashed lines. At sufficiently high temperatures the singularity instead bulges down.}
    \label{fig:PenroseDiagram}
    \vskip -0.1in
\end{figure}

An ingoing radial null geodesic that leaves the boundary at boundary time $t=0$ reaches the singularity at a value of the interior spatial coordinate $t_\star$ given by
\be\label{eq:tstar}
t_\star = PV \int_{0}^\infty \frac{e^{\chi(z)/2}}{f(z)}\,\mathrm{d}z\,,
\ee
where $PV$ denotes taking the principal value upon crossing the horizon at $z = z_{\mathcal{H}}$. The principal value indicates that the interior spatial $t$ coordinate is naturally related to the boundary time by a constant imaginary shift from the residue $-i \pi e^{\chi(z_{\mathcal{H}})/2}/ f'(z_{\mathcal{H}}) = i/(4T)$, but this shift is unnecessary to understand the bulk Penrose diagram of a purely real spacetime. Recall that $f>0$ outside the horizon and $f<0$ inside the horizon, so the integral in (\ref{eq:tstar}) could have either sign. The direction in which the singularity bends depends on the sign of $t_\star$. This is because 
in the black hole interior $t = 0$ corresponds to the midpoint of the Penrose diagram. If $t_\star > 0$, for example, then the geodesic has reached the singularity before reaching the midpoint of the diagram, and hence the singularity must have bent down. Similarly, if $t_\star < 0$ then the singularity bends up. Both of these possibilities are realized in our solutions and are shown in Fig.~\ref{fig:PenroseDiagram}.

We find that $t_\star \to -\infty$ both as we turn off the deformation, $\phi_{(0)}/\mu \to 0$, and also at low temperatures, $T/\mu \to 0$. In these limits the singularity therefore bends up and becomes null. At higher temperatures and larger deformations, the singularity  bends down. These two regimes are shown in Fig. \ref{fig:phaseDiag}.  The singularity can bend down more than that of Schwarzschild AdS, but approaches this in the limit $T/\mu \to \infty$ at fixed $\phi_{(0)}/\mu \ne 0$. Even though the limiting Penrose diagram resembles Schwarzschild AdS, the Kasner exponents are different, since $p_t \to 1$.\footnote{It may seem strange that the singularity does not become null as $p_t \to 1$, but there is no contradiction. When $p_t = 1$, the Kasner singularity becomes a smooth null surface, but for $p_t < 1$, the value of $t_*$ depends on global properties of the solution.} When $\phi_{(0)}=0 $, and the scalar condenses spontaneously, the singularity becomes null both as $T\to 0$ and $T \to T_c$.

\section{Dilatonic theories: Lifshitz to Kasner}

Einstein-Maxwell-dilaton theories have exact black hole solutions with no inner horizons and with Kasner singularities determined by parameters in the action \cite{Garfinkle:1990qj}. In this section we describe how these fixed exponents relate to the source-dependent Kasner exponents we have described so far. We will see that, in a holographic context,  the known explicit solutions describe a low temperature, near-horizon limit of a class of  geometries with  more general Kasner exponents.  The fixed exponents arise in this limit  in a sort of `attractor mechanism'. In direct analogy to the nondilatonic case, deformations away from this limit change the exponents by an amount that depends on the deformation. 

The simplest holographic setting for the physics we are after is the theory \cite{Taylor:2008tg, Goldstein:2009cv}
\be\label{eq:action2}
\Lag = R + 6  - \frac{e^{\g \phi}}{4} F^2 - g^{ab} \pa_a \phi \pa_b \phi \,.
\ee
There is a single coupling $\gamma$ in the Lagrangian.
With the same ansatz for the fields as we have been considering all along, the equations of motion are now
\begin{subequations}
\begin{align}
z^2 e^{\chi/2} \left(e^{-\chi/2} z^{-2} f \phi' \right)' & = - {\textstyle{\frac{1}{4}}} \g z^4 \rho^2 e^{-\g \phi} \,, \\
2 z^4 (z^{-3} f)' & = z^2 f (\phi')^2 - 6 + \half z^4 \rho^2 e^{-\g \phi} \,, \\
\chi' & = z (\phi')^2 \,,
\end{align}
\end{subequations}
and the electric field is
\be
\Phi' = - \rho \, e^{-\g \phi - \chi/2} \,.
\ee
These equations have an exact black hole solution given by \cite{Taylor:2008tg}
\be
f_\text{BH} = f_o \left(1 - (z/z_{\mathcal{H}})^{3 + 8/\g^2} \right) \,, \quad \chi_\text{BH} = \frac{16}{\g^2} \log z \,, \quad \phi_\text{BH} = \frac{4}{\g} \log z + \phi_o \,,
\ee
with the constants
\be
f_o = \frac{3 \g^4}{(4+\g^2) (8+3 \g^2)} \,, \qquad \phi_o = \frac{1}{\g} \log \frac{\rho^2 (4 + \g^2)}{48} \,.
\ee
This solution has the following asymptotics. As $z \to 0$ it tends to a so-called Lifshitz geometry with dynamical scaling exponent
\be
z_L = 1 + \frac{8}{\g^2} \,.
\ee
As $z \to \infty$ it tends towards a Kasner singularity with
\be\label{eq:kass}
p_t(\g) = \frac{8 - \g^2}{8 + 3 \g^2} \,, 
\ee
Thus in this solution the Kasner exponent is fixed by the parameter $\gamma$ in the theory. Despite the presence of a Maxwell field, these solutions are best thought of as a one-parameter family generalization of the Schwarzschild-AdS solution, which is recovered in the limit $\g \to \infty$ (wherein $z_L \to 1$ and $p_t \to -1/3$).

The exact solution discussed above arises as the near-horizon geometry of a near extremal black hole in an asymptotically AdS spacetime \cite{Goldstein:2009cv}. While not strictly necessary, it is clarifying to take this bigger perspective, in which case the geometry is divided into three regions for near-extremal states with $T \ll \mu$ (here $\mu$ is some UV energy scale, such as the chemical potential, that sets the crossover from AdS to Lifshitz):
\be\label{eq:threeregions}
\text{AdS UV} \; \xrightarrow{z \gg \mu^{-1}} \; \text{Lifshitz IR: } z_L \; \xrightarrow{z \gg T^{-1/z_L}} \; \text{Kasner transhorizon: } p_t \,.
\ee
The important point here is the following. If the Lifshitz geometry is obtained in this way by flowing from some AdS UV, then the solution cannot be pure Lifshitz at any finite $z$. There must also be irrelevant deformations that decay as $z \to \infty$. These are the deformations that flow the Lifshitz solution back up to AdS.
At any nonzero temperature, where $z_{\mathcal{H}}$ is finite, these deformations will be nonzero on the horizon. We will see that these deformations on the horizon shift the Kasner exponent away from the value $p_t(\g)$ in (\ref{eq:kass}). As $T \to 0$, $z_{\mathcal{H}} \to \infty$ and the deformation becomes small on the horizon so that $p_t \to p_t(\g)$. This is a transhorizon manifestation of the attractor mechanism --- really just an IR fixed point in the RG sense --- discussed in \cite{Goldstein:2009cv}. However, at any $T > 0$ the value of $p_t$ is different and depends on the strength of the irrelevant deformation.

The simplest point is to verify that Kasner scalings with more general exponents than (\ref{eq:kass}) are consistent asymptotic near-singularity behaviors of the theory. The only constraint on the asymptotic Kasner exponent from the equations of motion is that
\be
p_t > - \frac{\g}{\g + 2 \sqrt{2 + \g^2}} \,.
\ee
At $\g=0$ this recovers the constraint that $p_t > 0$ that we found in (\ref{eq:pt}) for Einstein-Maxwell theory. As $\g \to \infty$, the lower bound goes down to the Schwarzschild value of $-1/3$, which is also consistent.

To see explicitly how a source shifts the Kasner exponent it is sufficient to work within the Lifshitz IR scaling regime. The irrelevant deformation appears as a source $\delta \phi_{(0)}$ for the scalar field at the Lifshitz boundary (because the mode will be irrelevant and grow towards the UV, the source should be imposed at some small but nonzero cutoff $z_\text{UV}$). This will lead to a linearized perturbation of the bulk fields about the Lifshitz black hole background. Radial perturbations are easily seen to have the general form
\be
\phi = \phi_\text{BH} + \delta \phi \,, \quad \chi = \chi_{BH} + \frac{8}{\g} \delta \phi + \delta \chi_o \,, \quad f = f_\text{BH}\left(1 + \frac{4}{\g} \delta \phi \right) + \delta f_o (z/z_{\mathcal{H}})^{3+8/\g^2}\,, \label{eq:shift}
\ee
where $\delta \chi_o$ and $\delta f_o$ are constants and $\delta \phi$ must obey
\be\label{eq:phieq}
z^4 e^{\chi_\text{BH}/2} \left(z^{-2} e^{-\chi_\text{BH}/2} f_\text{BH} \, \delta \phi '\right)' = 12 \, \delta \phi \,.
\ee
We will focuss on $\delta \phi$. The constants $\delta \chi_o$ and $\delta f_o$ can be chosen to keep either the energy density (sourced by $\delta g_{tt}$) or the temperature constant as we deform by the scalar operator. This choice does not affect the considerations below.

The scalar equation (\ref{eq:phieq}) can be solved in terms of Gaussian hypergeometric functions. The solution to this equation that is regular on the horizon takes the form
\be
\delta \phi = c_\Delta \left(\frac{z}{z_{\mathcal{H}}}\right)^{\Delta}  {}_2F_{1}\left(\frac{\Delta}{2+z_L},\frac{\Delta}{2+z_L};\frac{2\Delta}{2+z_L};\left(\frac{z}{z_{\mathcal{H}}}\right)^{2 + z_L} \right)
- \Big(\Delta \leftrightarrow 2 + z_L - \Delta \Big) \,,
\ee
where the scaling dimension is $\Delta = [8 + 3 \g^2 + \sqrt{(8+3\g^2)(72 + 19 \g^2)}]/{2 \g^2}$ and
the coefficient is $c_\Delta = \varepsilon \, \Gamma\left( \frac{ \Delta}{2 + z_L} \right)^2 \Big/\, \Gamma\left( \frac{2 \Delta}{2 + z_L} \right)$. The small number $\varepsilon$ can be related to the source $\delta \phi_{(0)}$ by expanding near the Lifshitz boundary as $z \to 0$, where
$\delta \phi = \delta \phi_{(0)} z^{2+z_L - \Delta} +  \delta \phi_{(1)} z^\Delta + \cdots$.
Clearly $\vep \propto \delta \phi_{(0)}$. One immediately verifies that $2+z_L - \Delta < 0$ for all $\g$, so that $\delta \phi_{(0)}$ is indeed an irrelevant deformation of the Lifshitz fixed point as we expected.

Expanding the solution beyond the horizon as $z \to \infty$ we find
\be
\delta \phi = 2 \vep \cos \frac{\pi \Delta}{2 + z_L} \cdot \log z^{2 + z_L} + \cdots \,.
\ee
This logarithmic growth towards the singularity amounts to a linearized shift in the Kasner exponent to the value
\be\label{eq:dilkas}
p_t = p_t(\g) + \frac{32 \g \, \varepsilon}{8 + 3 \g^2}  \cos \frac{\pi \Delta \g^2}{8 + 3 \g^2} \,.
\ee
The strength $\varepsilon$ is given, on dimensional grounds, in terms of the strength $\delta \phi_{(0)}$ of the deformation and the temperature $T$ as
\be\label{eq:shift2}
\delta p_t \; \propto \; \varepsilon \; \propto \delta \phi_{(0)} T^{(\Delta - 2 - z_L)/z_L} \,.
\ee
For this irrelevant deformation $\Delta > 2 + z_L$. Therefore, as we should expect, the shift becomes small as the temperature goes to zero (and hence the perturbative computation is self-consistent in this limit). That is because the perturbation decays towards the IR in the Lifshitz region outside the horizon. As the horizon goes deeper into the IR, the perturbation on the horizon becomes smaller. Once past the horizon, the perturbation starts to grow logarithmically and shifts the Kasner exponent. This shift of the exponent is therefore smaller at small temperatures as we see in (\ref{eq:shift2}). The value $p_t(\g)$ is achieved in the limit $T \to 0$.

\section{Traversing geodesics}

The Penrose diagram in Fig. \ref{fig:PenroseDiagram} has 
two boundaries corresponding, as usual, to the two copies of the dual field theory that have been entangled in a thermofield double state. A natural set of boundary observables are correlations functions of large dimension operators between the two copies. These are described in the bulk by spacelike geodesics that traverse the Einstein-Rosen bridge, going from one boundary to the other. The information contained in such Schwinger-Keldysh correlation functions can also be obtained from the retarded Green's function, that depends solely on the black hole exterior. See \emph{e.g.} \cite{Herzog:2002pc} for a holographic discussion. Nonetheless, the transhorizon perspective can reveal interesting features of these correlation functions in a transparent way. In particular, we will now see that our charged black holes all have a purely decaying `overdamped' quasinormal mode that can be related to a maximum of $g_{tt}$ in the black hole interior. The existence of this maximum can be thought of as a remnant of the (now absent) Reissner-Nordstr\"om Cauchy horizon.

Radial spacelike geodesics in the black hole background can be labelled by a constant `energy' $E$. These geodesics fall into the black hole up to a turning point $z_\star$ given by \cite{Fidkowski:2003nf,Frenkel:2020ysx}:
\be
E^2 = g_{tt}(z_\star) \,.\label{eq:E2z}
\ee
Recall that $g_{tt} =  - f e^{-\chi}/z_\star^2 > 0$
beyond the horizon. After reaching the turning point, the geodesics emerge on the other side of the Einstein-Rosen bridge. The behavior of the geodesic in the interior depends upon the form of $g_{tt}(z)$. Clearly $g_{tt}$ vanishes on the horizon. If $g_{tt}$ increases without bound beyond the horizon then $z_\star \to \infty$ as $E \to \infty$. These geodesics can come arbitrarily close to the singularity \cite{Fidkowski:2003nf}. However, if $g_{tt}$ has a maximum at some $z_\text{c}$ beyond the horizon, i.e. with $g_{tt}'(z_\text{c}) = 0$, then geodesics anchored at the boundary get `stuck' at this critical value and do not come closer to the singularity \cite{Hartman:2013qma}.

In the asymptotic Kasner regime $g_{tt} \sim z^{1-\a^2} \sim z^{-4 p_t/(1-p_t)}$. If $p_t > 0$ then $g_{tt} \to 0$ asymptotically, while if $p_t < 0$ then $g_{tt} \to \infty$. Our charged black holes have $p_t > 0$, and therefore $g_{tt}$ must have a maximum at some intermediate $z_\text{c}$. This maximum is visible, for example, in Fig. \ref{fig:collapse} above. In contrast, neutral black holes deformed by a scalar field source necessarily have $p_t < 0$ and there is no critical radius for real geodesics \cite{Frenkel:2020ysx}.

It was explained in \cite{Hartman:2013qma} that if
real (as opposed to complex) geodesics get stuck at a critical interior radius $z_\text{c}$, then large mass scalar fields in the black hole exterior have an overdamped, non-oscillatory, quasinormal mode. The mode decays as $e^{- \Gamma t}$ with decay rate $\Gamma$ determined directly from the black hole interior as 
\be\label{eq:mode}
\Gamma = M \sqrt{g_{tt}(z_\text{c})}  \,.
\ee
Here $M$ is the large mass of the scalar field. We have verified the existence of this precise mode directly from numerical computation of perturbations in the black hole exterior. More general, oscillating, quasinormal modes are instead related to complex geodesics \cite{Motl:2003cd, Fidkowski:2003nf, Festuccia:2005pi, Festuccia:2008zx, Hartman:2013qma}.

In our solutions the maximum of $g_{tt}$ is in between the horizon and the would-be inner horizon, where the ER bridge collapses. Indeed, the maximum exists also for RN-AdS, where $g_{tt}$ vanishes at both horizons and must therefore have a maximum in between. In this sense, we can think of the existence of this maximum (and hence the overdamped mode (\ref{eq:mode})) in our solutions as a remnant of the RN-AdS inner horizon. It is not obvious a priori that the maximum would survive with large boundary deformations, but the fact that $p_t > 0$ implies that it does.

\section{Discussion}

We have studied the gravitational dual of the grand canonical ensemble of a CFT deformed by relevant or irrelevant operators. These black hole spacetimes are more generic than the familiar Reissner-Nordstr\"om AdS solution, which is the most widely studied dual to the grand canonical ensemble of a CFT. The region of spacetime inside the horizon turns out to be quite different, and has some interesting properties. We have shown that Cauchy horizons never arise for relevant perturbations dual to a bulk scalar with $m^2 < 0$ (but above the BF bound). Instead, the spacetime ends in a spacelike Kasner singularity. For small deformations, the Kasner phase is preceded by a dramatic collapse of the Einstein-Rosen bridge connecting the two asymptotic regions.

It remains an open question how the experience of an infalling observer is encoded in the dual field theory. Even though we expect the classical description of such an observer to break down near the singularity, we can ensure that quantum and stringy corrections remain small until we are well within the Kasner epoch by taking large $N$ and large coupling in the field theory.\footnote{Since we have not compactified any directions,  classical stringy effects like winding modes becoming tachyonic do not occur.} By constructing more generic black hole interiors, as we have done, we can start to understand the classical data that is needed to characterize the approach to the singularity. This data --- such as the Kasner exponents --- must be part of any eventual field theoretic understanding of the fate of infalling observers or of the black hole interior more generally.

We conclude with a few comments that extend some of our results. Firstly, we describe a different setting in which Cauchy horizons can survive scalar field deformations at fine-tuned values of the parameters. We have seen that Cauchy horizons can exist at a certain discrete set of $m^2 > 0$ for each temperature. Without changing the scalar potential, these are specific irrelevant deformations which will destroy the asymptotic AdS boundary. It is interesting to note that one can also construct asymptotically AdS solutions with a smooth Cauchy horizon and a simple quadratic potential with negative mass squared, $m^2 < 0$. This can be achieved with two complex scalar fields $\phi_1, \phi_2$ and a (slightly) inhomogeneous field configuration. Consider the theory (\ref{eq:action}) with two complex scalars with the same $m^2 < 0$. Suppose
\be
\phi_1 = \phi(z) e^{i k x}, \qquad \phi_2 = \phi(z) e^{i k y} \,.
\ee
The stress tensor and hence the metric then remain homogeneous and isotropic. These are examples of holographic Q-lattices \cite{Donos:2013eha}. By a similar analysis as in \S3.2, at the linearized level the condition of a regular Cauchy horizon translates into an eigenvalue problem for $k^2$  with a discrete set of solutions. Each of these solutions can then be extended to a full nonlinear solution. These solutions do not violate strong cosmic censorship since they are still nongeneric, but it is interesting that there are simple  deformations of the dual CFT that do not decay in time and still preserve a Cauchy horizon. If one allows boundary sources that are unbounded, there are even simpler examples. If $m=0$ and $\phi_1=a\,x$, $\phi_2=a\,y$, there is a particularly well studied isotropic black brane \cite{Bardoux:2012aw,Iizuka:2012wt,Andrade:2013gsa}, where
\begin{equation}
    f(z)=1-a^2 z^2+\frac{z^4 \rho ^2}{4}-\left(\frac{z}{z_{\mathcal{H}}}\right)^3 \left(1-a^2 z_{\mathcal{H}}^2+\frac{z_{\mathcal{H}}^4 \rho ^2}{4}\right) \,,
\end{equation}
with $\chi=0$ and $\Phi$ as in Eq.~(\ref{eq:E}). So long as $\frac{\rho ^2 z_{\mathcal{H}}^4}{4}+a^2 z_{\mathcal{H}}^2\leq 3$ and $\rho>0$ a smooth Cauchy horizon exists in the interior of the black hole\footnote{The fine tuned case with $\rho=0$ is more intricate. For $a^2 z_{\mathcal{H}}^2<1$, there is no Cauchy horizon and the interior looks similar to a Schwarzschild black brane, with $p_t = -1/3$. For $1< a^2 z_{\mathcal{H}}^2<3$ there is a smooth Cauchy horizon, with the upper bound $a^2 z_{\mathcal{H}}^2=3$ representing a smooth extremal black brane. For $a^2 z_{\mathcal{H}}^2=1$ there is no Cauchy horizon, and $p_t = 0$.}.

Secondly, it is straightforward to generalize our analysis to $d+1$ bulk spacetime dimensions. The equations of motion become
\begin{subequations}
\begin{align}
z^{d-1}e^{\chi/2}(z^{1-d}f e^{-\chi/2}\phi')' &= \frac{m^2}{z^2}\phi,\\
2(d-1)z^{d+1} (z^{-d} f)' &= - 2d(d-1) + 2 m^2 \phi^2 + z^4 e^{\chi}(\Phi')^2 + 2 z^2 f (\phi')^2 \,, \\
\chi' &= \frac{2 z}{d-1} (\phi')^2 \,,
\end{align}
\end{subequations}
and we can again solve the Maxwell equation explicitly: $\Phi' = - \rho z^{d-3}e^{-\chi/2}$. All our results still go through. There is still rapid collapse of the ER bridge for small deformations and at large $z$ the geometry has a Kasner behaviour,
\be 
f = - f_o z^{\frac{2\a^2}{d-1} + d},\quad \phi = \a \sqrt{2} \log z, \quad \chi = 4 \frac{\a^2}{d-1}\log z + \chi_o,
\ee
with $\a^2 > (d-2)(d-1)/2$. The Kasner exponent
\be 
p_t = \frac{2\a^2 - (d-1)(d-2)}{2\a^2 + d(d-1)},\quad 1 \geq p_t > 0.
\ee
Furthermore, it is straightforward to check that the proof of no Cauchy horizons discussed in \S \ref{sec:proofnoC} goes through for general dimensions.

\subsection*{Acknowledgments}

We thank Roberto Emparan, Harvey Reall and Eva Silverstein for insightful discussions. S.~A.~H. is supported by DOE award DE-SC0018134 and by a Simons Investigator award.
G.~H. is supported in part by NSF grant PHY1801805.  J.~K. is supported by the Simons foundation. J.~E.~S. is supported in part by STFC grants PHY-1504541 and ST/P000681/1. J.~E.~S. also acknowledges support from a J.~Robert~Oppenheimer Visiting Professorship. This work used the DIRAC Shared Memory Processing system at the University of Cambridge, operated by the COSMOS Project at the Department of Applied Mathematics and Theoretical Physics on behalf of the STFC DiRAC HPC Facility (www.dirac.ac.uk). This equipment was funded by BIS National E- infrastructure capital grant ST/J005673/1, STFC capital grant ST/H008586/1, and STFC DiRAC Operations grant ST/K00333X/1. DiRAC is part of the National e-Infrastructure. 

\providecommand{\href}[2]{#2}\begingroup\raggedright\endgroup

\end{document}